# Universal Equation of States are Derived from the Isothermal Relationships of Elastic Solids

József Garai

*Department of Mechanical and Materials Engineering, Florida International University, Miami, FL 33174,*

*USA*

Universal P-V-T equations of states are derived from the original isothermal expressions of Birch-Murnaghan and the Vinet EoSs by using absolute reference frame and incorporating the temperature dependence of the parameters. The equations are tested against the experiments of perovskite [MgSiO$_3$] and periclase [MgO]. The determined parameters of the EoSs are valid throughout the entire pressure and temperature ranges, which cover 0-142 GPa and 293-3000 K and 0-109 GPa and 293-2199 K for periclase and perovskite respectively. The root-mean-square (RMS) misfits of the universal Birch-Murnaghan EoS are 0.73 and 0.33 GPa and the misfit of the Vinet EoS are 0.74 and 0.32 GPa for perovskite and periclase respectively.

The pressure-volume (P-V) relationship of solids is described by isothermal equation of states. The most widely used isothermal EoSs in solid state physics are the Birch-Murnaghan and the Vinet. The third-order Birch-Murnaghan EoS [1-3] is given as:

$$p = \frac{3B_{0T}}{2}\left[\left(\frac{V_{0T}}{V}\right)^{\frac{7}{3}} - \left(\frac{V_{0T}}{V}\right)^{\frac{5}{3}}\right]\left\{1 + \frac{3}{4}(B_0' - 4)\left[\left(\frac{V_{0T}}{V}\right)^{\frac{2}{3}} - 1\right]\right\}, \qquad (1)$$

where $B_{0T}$, $B_0'$ and $V_{0T}$ are the bulk modulus, the pressure derivative of the bulk modulus and volume respectively at zero pressure and at the temperature of interest. The so-called universal EoS derived by Rose [4] from a general inter-atomic potential and promoted by Vinet [5] is:

$$p = 3B_{0T}\frac{1-f_V}{f_V^2}e^{\left[\frac{3}{2}(B_0'-1)(1-f_V)\right]} \qquad (2)$$

where

$$f_V = \left(\frac{V}{V_{0T}}\right)^{\frac{1}{3}}. \tag{3}$$

Simple approach for describing the P-V-T relationship of elastic solids is to raise the temperature of the material first and then use the isothermal EoS [e.g. 6]. In order to use the isothermal EoS the parameters $B_{0T}$, $B_0'$ and $V_{0T}$ must be known at the temperature of interest. The bulk modulus and volume at the temperature of interest can be calculated as:

$$B_{T0}(T) = B_{T0}(T_0) + \left(\frac{\partial B_T}{\partial T}\right)_P (T - T_0). \tag{4}$$

and

$$V_0(T) = V_0(T_0) e^{\int_{T_0}^{T} \alpha(T) dT} \tag{5}$$

where $\alpha(T)$ is the volume coefficient of thermal expansion at ambient-pressure [e.g. 6, 7]. The temperature dependence of the volume coefficient of thermal expansion can be described as:

$$\alpha(T) = a + bT - \frac{c}{T^2}. \tag{6}$$

Substituting Eqs. (4)-(6) into the isothermal EoSs allows describing the P-V-T relationship of solids within a limited temperature range. Usually one set of parameters is capable to cover a temperature range of 500-800 K [7]. Beyond this range new sets of parameters are needed for an accurate description. Many of the parameters in the EoS are inter-related, which adds to the complexity of calculations. The optimum values of each of the interrelated parameters have to be determined by confidence ellipses [e.g. 7, 8]. The thermodynamic description of solids covering wide temperature and pressure ranges are complicated and time consuming.

In this study a universal form is proposed for the isothermal EoSs. It is suggested that absolute reference frame should be used for the universal description. The zero pressure and temperature initial parameters, volume $[V_o]$, bulk modulus $[B_o]$ and volume coefficient of thermal expansion $[\alpha_o]$ can be defined as:

$$V_o \equiv nV_o^m \tag{7}$$

and

$$B_o \equiv \lim_{p \Rightarrow 0} B_{T=0} \tag{8}$$

and

$$\alpha_o \equiv \lim_{T \Rightarrow 0} \alpha_{Vp=0} \tag{9}$$

respectively, where $V_o^m$ is the molar volume at zero pressure and temperature and n is the number of moles. The volume at zero pressure and at a given temperature can be calculated as:

$$V_{0T} = V_o e^{\int_{T=0}^{T} \alpha_T dT} . \tag{10}$$

Assuming that the temperature effect on volume coefficient of thermal expansion is linear [c=0 in Eq. (6)] it can be written as:

$$\alpha_T \cong \alpha_o + \alpha_{01} T . \tag{11}$$

Approximate sign is used for the temperature dependence of the volume coefficient of thermal expansion in Eq. (11) because the temperature dependence of the coefficient below the Debye temperature is not linear but rather correlates to the heat capacity [9]. The fitting results of the two substances used in this study indicate that the introduced error is minor and Eq. (11) can be used for P-V-T calculations. Substituting Eq. (11) into Eq. (10) gives the volume at temperature T

$$V_{0T} \cong V_o e^{\int_{T=0}^{T} (\alpha_0 + \alpha_{01} T) dT} \cong V_o e^{(\alpha_o + \alpha_1 T)T} . \tag{12}$$

By assuming that the product of the volume coefficient of thermal expansion and the bulk modulus is constant the temperature dependence of the bulk modulus at 1 bar pressure can be derived from fundamental thermodynamic relationships [10]

$$B_{0T} = B_o e^{-\int_{T=0}^{T} \alpha_T \delta dT} \tag{13}$$

where $\delta$ is the Anderson-Grüneisen parameter at ambient conditions. Assuming constant value for the Anderson-Grüneisen parameter, which is reasonable at temperatures higher than the Debye temperature [10] and substituting Eq. (11) into Eq. (13) gives the temperature dependence of the bulk modulus as:

$$B_{0T} = B_o e^{-\int_{T=0}^{T}(\alpha_0+\alpha_{01}T)\delta dT} \cong B_o e^{-(\alpha_0+\alpha_1 T)\delta T}. \qquad (14)$$

Assuming that the pressure derivative of the bulk modulus remains constant and plugging the thermal EoS [Eq. (12)] and the temperature dependence of the bulk modulus [Eq. (14)] into the original isothermal EoSs [Eqs. (1)-(3)] results in a universal $(V,T) \Rightarrow p$ description of solids. The universal form for Birch-Murnaghan and Vinet EoSs can be written then as:

$$p = \frac{3B_o e^{-(\alpha_0+\alpha_1 T)\delta T}}{2} \left[\left(\frac{V_o e^{(\alpha_0+\alpha_1 T)T}}{V}\right)^{\frac{7}{3}} - \left(\frac{V_o e^{(\alpha_0+\alpha_1 T)T}}{V}\right)^{\frac{5}{3}}\right]$$
$$\left\{1 + \frac{3}{4}(B_o' - 4)\left[\left(\frac{V_o e^{(\alpha_0+\alpha_1 T)T}}{V}\right)^{\frac{2}{3}} - 1\right]\right\} \qquad (15)$$

and

$$p = 3B_o e^{-(\alpha_0+\alpha_1 T)\delta T} \frac{1-\left(\frac{V}{V_o e^{(\alpha_0+\alpha_1 T)T}}\right)^{\frac{1}{3}}}{\left(\frac{V}{V_o e^{(\alpha_0+\alpha_1 T)T}}\right)^{\frac{2}{3}}} e^{\left\{\frac{3}{2}(B_o'-1)\left[1-\left(\frac{V}{V_o e^{(\alpha_0+\alpha_1 T)T}}\right)^{\frac{1}{3}}\right]\right\}} \qquad (16)$$

respectively.

Perovskite (MgSiO$_3$) and periclase (MgO) are the most abundant materials in the Earth's lower mantle [11]. This significant interest in geophysics results in the availability of experiments covering wide pressure and temperature range. The 269 experimental data of perovskite [12 and ref. therein] cover the pressure and the temperature range of 0-109 GPa and 293-2199 K. The experiments (360) of periclase are conducted between 0-142 GPa and 0-3000 K [13 and ref. therein].

The fitting accuracy of the EoSs is evaluated by RMSD and Akaike Information Criteria AIC [14, 15]. The Akaike Information Criteria is devised assessing the right level of complexity. Assuming normally distributed errors, the criterion is calculated as:

$$\text{AIC} = 2k + n \ln\left(\frac{\text{RSS}}{n}\right), \qquad (17)$$

where n is the number of observations, RSS is the residual sum of squares, and k is the number of parameters. AIC penalizes both for increasing the number of parameters and for reducing the size of data. The preferred model is the one which has the smallest AIC value.

The RMS misfits for the complete data set (360) of periclase are 0.54 GPa and 0.55 GPa for the Universal Birch-Murnaghan and Vinet EoSs respectively while the RMS misfits for perovskite (269) are 0.78 GPa and 0.79 GPa respectively.

The original data set was optimized by removing experiments with high misfit. Repeating the fitting for the new set of data the removal of the experiments was accepted as long as the AIC value of the new data set is smaller than the initial one. After removing 34 experiments and satisfying the AIC fitting criteria the RMS misfit of periclase improved to 0.33 GPa, 0.32 GPa for the Universal Birch-Murnaghan and Vinet EoSs respectively. Removing 5 experiments improved the RMS misfit of perovskite to 0.73 GPa, 0.74 GPa for the Universal Birch-Murnaghan and Vinet EoSs respectively. The excellent fitting of the data to the EoSs indicate that all three universal EoS correctly describe the P-V-T relationship of periclase and perovskite. The fitting parameters and results are given in TAB. I and II. The fitting parameters for the recently proposed universal EoS [12] are also listed. In order to compare the determined parameters to previous studies the ambient condition values were calculated [TAB. III]. The values are consistent with previous investigations [16, 17].

Based on the RMS misfit and AIC values the fittings of the conventional EoSs are slightly better than the EoS [12]. It should be noted that the conventional EoSs are minimizing the misfit to the pressure while the EoS [12] minimize the misfit to the volume. Thus higher misfit for the pressure is expected when the misfit is minimized originally to the volume.

Using absolute reference frame, zero pressure and temperature, and incorporating the temperature dependence of the volume and the bulk modulus new universal EoSs are derived from the isothermal Birch-Murnaghan and Vinet EoSs. These universal EoSs contain seven parameters such as initial molar volume, initial bulk modulus, initial volume coefficient of thermal expansion, number of moles, temperature derivative of the volume coefficient of thermal expansion, pressure derivative of the bulk modulus and the Anderson-Grüneisen parameter. These parameters remain constant throughout the entire pressure and temperature range and sufficient to describe the universal P-V-T relationship of elastic solids with high accuracy. The derived universal EoSs were tested against the experiments of perovskite and periclase with positive results. The RMS misfits of the EoSs are slightly higher than the uncertainty of the experiments.

**Acknowledgement**

The author thank to Sergio Speziale for reading and commenting the manuscript.

TABLE I. Fitting parameters and results of periclase for the total and optimized data sets.

| EoS | Number of experiments | $B_o$ [GPa] | $V_o$ [cm$^3$] | $B_o'$ | $\alpha_o$ x10$^{-5}$ | $\alpha_1$ x10$^{-9}$ | $\delta$ | a | b x10$^{-3}$ | c x10$^{-7}$ | d x10$^{-9}$ | Pressure RMS misfit | AIC |
|---|---|---|---|---|---|---|---|---|---|---|---|---|---|
| Universal $(V,T) \Rightarrow p$ | 360 | 161.56 | 11.153 | 4.135 | 2.909 | 6.835 | 3.597 | | | | | 0.535 | -438.1 |
| Birch-Murnaghan (Eq. 15) | 324 | 167.62 | 11.119 | 4.131 | 3.522 | 4.751 | 4.893 | | | | | 0.329 | -708.0 |
| Universal $(V,T) \Rightarrow p$ | 360 | 157.41 | 11.161 | 4.488 | 2.892 | 6.770 | 3.572 | | | | | 0.547 | -422.3 |
| Vinet (Eq. 16) | 324 | 160.07 | 11.145 | 4.474 | 3.106 | 6.165 | 3.645 | | | | | 0.318 | -729.6 |
| EoS [Garai, 2007] | 360 | 169.14 | 11.131 | | 2.956 | | | 1.681 | -2.195 | -1.858 | 7.220 | 0.678 | -266.2 |
| | 324 | 166.72 | 11.136 | | 3.029 | | | 1.693 | -2.172 | -1.931 | 6.698 | 0.352 | -676.2 |

TABLE II. Fitting parameters and results of perovskite for the total and optimized data sets.

| EoS | Number of data | $B_o$ [GPa] | $V_o$ [cm$^3$] | $B_o'$ | $\alpha_o$ x10$^{-5}$ | $\alpha_1$ x10$^{-9}$ | $\delta$ | a | b x10$^{-3}$ | c x10$^{-7}$ | d x10$^{-9}$ | f | Pressure RMS misfit | AIC |
|---|---|---|---|---|---|---|---|---|---|---|---|---|---|---|
| Universal $(V,T) \Rightarrow p$ | 269 | 260.80 | 24.251 | 4.016 | 2.591 | -1.089 | 3.149 | | | | | | 0.784 | -118.7 |
| Birch-Murnaghan (Eq.15) | 265 | 260.79 | 24.252 | 4.060 | 2.576 | -0.802 | 3.355 | | | | | | 0.735 | -151.3 |
| Universal $(V,T) \Rightarrow p$ | 269 | 258.37 | 24.253 | 4.214 | 2.619 | -1.247 | 3.119 | | | | | | 0.787 | -116.7 |
| Vinet (Eq. 16) | 265 | 258.23 | 24.253 | 4.267 | 2.609 | -0.976 | 3.325 | | | | | | 0.738 | -149.2 |
| EoS [Garai, 2007] | 269 | 267.51 | 24.284 | | 2.079 | | | 1.556 | 0 | -1.098 | 0 | | 0.792 | -115.6 |
| | 269 | 273.26 | 24.304 | | 1.616 | | | 1.392 | 0.900 | -0.585 | 4.301 | 10.23 | 0.786 | -113.3 |
| | 265 | 272.55 | 24.303 | | 1.647 | | | 1.424 | 0.884 | -0.711 | 4.040 | 7.86 | 0.738 | -145.9 |

TABLE III. Parameters at ambient condition.

| | EoS | $V_0$ [cm$^3$] | $K_0$ [GPa] | $B_o'$ | $\alpha_o$ x10$^{-5}$ |
|---|---|---|---|---|---|
| Perovskite (MgSiO3) | Universal $(V, T) \Rightarrow p$ Birch-Murnaghan (Eq.15) | 24.437 | 254.24 | 4.060 | 2.552 |
| | Universal $(V, T) \Rightarrow p$ Vinet (Eq. 16) | 24.440 | 251.73 | 4.267 | 2.580 |
| Periclase (MgO) | Universal $(V, T) \Rightarrow p$ Birch-Murnaghan (Eq.15) | 11.251 | 158.93 | 4.131 | 3.663 |
| | Universal $(V, T) \Rightarrow p$ Vinet (Eq. 16) | 11.524 | 154.47 | 4.474 | 3.298 |